# Physical Layer Network Coding with Multiple Antennas

Shengli Zhang[†‡]    Soung Chang Liew[†]

[†] Department of Information Engineering, the Chinese University of Hong Kong, Hong Kong

[‡] Department of Communication Engineering, Shenzhen University, China. {slzhang, soung}@ie.cuhk.edu.hk

**Abstract:** The two-phase MIMO NC (network coding) scheme can be used to boost the throughput in a two-way relay channel in which nodes are equipped with multiple antennas. The obvious strategy is for the relay node to extract the individual packets from the two end nodes and mix the two packets to form a network-coded packet. In this paper, we propose a new scheme called MIMO PNC (physical network coding), in which the relay extracts the summation and difference of the two end packets and then converts them to the network-coded form. MIMO PNC is a natural combination of the single-antenna PNC scheme and the linear MIMO detection scheme. The advantages of MIMO PNC are many. First, it removes the stringent carrier-phase requirement in single-antenna PNC. Second, it is linear in complexity with respect to the constellation size and the number of simultaneous data streams in MIMO. Simulation shows that MIMO PNC outperforms the straightforward MIMO NC significantly under random Rayleigh fading channel. Based on our analysis, we further conjecture that MIMO PNC outperforms MIMO NC under all possible realizations of the channel.

## I. INTRODUCION

In wireless networks, the use of relay has many advantages. It can lead to better coverage and connectivity. With a smaller distance for node-to-node transmissions, the power consumption can be reduced. At the same time, the detrimental effects of the interferences from other transmissions can be alleviated, leading to higher capacity per unit area.

Consider the simple two-way relay channel (TWRC) shown in Fig. 1. In [1], the authors introduced network coding into TWRC: the two end nodes transmit their packets to the relay in two different time slots; the relay then forms a network-coded packet out of the two packets and broadcast it to the end nodes. The number of time slots needed to exchange one packet is 3. Subsequent to [1], we proposed physical layer network coding (PNC) [2]. PNC allows the two end nodes to transmit their packets in the same time slot. The superimposed packets received simultaneously are then directly transformed to a network-coded packet at the physical layer of the relay. As a result, the number of time slots needed to exchange one packet is reduced to 2.

PNC is attracting increasing attention. At the communication level, variants of PNC have been proposed [3, 4, 5] to improve performance or to ease implementation. At the network level, PNC has also been shown to be able to increase network capacity by a fixed factor [6, 7]. In addition, information-theoretic studies indicate that PNC can allow the capacity of TWRC to be approached in both low SNR and high SNR regions [8, 9, 10].

To date, most work on PNC assumes single antenna at the wireless devices. Since multiple-input-multiple-output (MIMO) can increase the channel capacity, and multiple antennas have been widely equipped in most modern wireless devices, the combination of PNC with MIMO will be of great interest. To the authors' knowledge, little work has been done on this front.

Refs. [11, 12] explored this combination, assuming the availability of full channel state information (CSI) at the two transmitting nodes (end nodes). The end nodes exploit the CSI to pre-code the packet before transmission. The pre-coding essentially multiplies the inverse of the channel matrix to the MIMO inputs before transmission. This cancels out the effect of the MIMO channel. This pre-equalization, however, requires the packets of the two end nodes to be synchronized (including carrier-phase synchronization) when they arrive at the relay. This imposes a significant implementation difficulty. The maximum likelihood (ML) based detection and encoding schemes in [5] can also be extended to the MIMO case without the need for carrier phase synchronization. However, the complexity increases exponentially with the constellation size and the number of data streams transmitted simultaneously from the end nodes.

In this paper, we propose a new MIMO PNC scheme in which the relay extracts the summation and difference of the two end packets and then converts them to the network-coded form. Our scheme only requires CSI only at the receiver. It can be regarded as a natural extension of the single-antenna PNC [2], and its advantages are also similar. A significant implication, however, is that unlike the single-antenna PNC, our MIMO PNC scheme gets rid of the requirement for carrier-phase synchronization, bringing implementation closer to reality. Also significant is the fact that instead of the exponential complexity in [5], our scheme, which makes use of linear MIMO detection methods, is linear in complexity.

For comparison purposes, this paper also considers the two-phase MIMO NC scheme in which MIMO technique is used to extract the individual packets from the two end nodes before converting them into a network-coded packet (as opposed to our MIMO_PNC in which the overlapped packets from the two end nodes are directly converted into a network-coded packet without extracting the individual packets). Analysis and simulation results show that MIMO PNC can achieve much better BER performance than MIMO NC.

The rest of this paper is organized as follows. Section II defines the system model and illustrates the basic idea of MIMO PNC with an example. Section III presents the details MIMO PNC, assuming two antennas at the relay and one antenna at the end nodes. The BER performance is analyzed. Section IV provides numerical simulation results that demonstrate the superiority of MIMO PNC over other schemes. This is followed by a discussion of the general MIMO PNC setting in which the two end nodes are also equipped with multiple antennas. Section V concludes this paper.

## II. SYSTEM MODEL AND ILLUSTRATING EXAMPLE

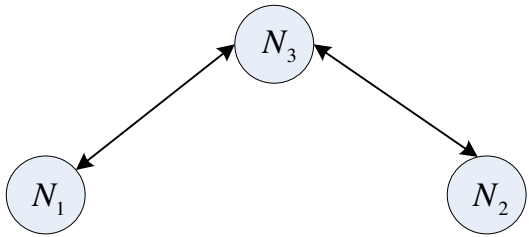

Figure 1. Two way relay channel

**A. System Model:**

This paper considers the two-way relay channel as shown in Fig. 1. The two end nodes, $N_1$ and $N_2$, exchange information through the relay node $N_3$. There is no direct link between the two end nodes. For simplicity, we assume the end nodes are equipped with single antenna and the relay node is equipped with two antennas.

The PNC transmission consists of two phase. In the first phase, both end nodes transmit to the relay node simultaneously. Here, we assume the two end nodes' signals arrive at the relay node at a symbol level synchronization. Then, the received signal at the relay node can be expressed as:

$$r_1 = h_{11}x_1 + h_{12}x_2 + n_1$$
$$r_2 = h_{21}x_1 + h_{22}x_2 + n_2 \qquad (1)$$

where $r_i$ denotes the received baseband signal at the $i$-th antenna of the relay node, $h_{i,j}$ is the complex Gaussian channel coefficient from node $N_j$ to the $i$-th antenna of the relay node, $x_i$ is the transmitted baseband signal of node $N_i$, and $n_j$ is the Complex Gaussian noise at the $j$-th antenna of the relay node with zero mean and variance $\sigma^2$ for both dimensions. BPSK modulation is assumed at both nodes (all the schemes presented in this paper can be easily extended to QPSK, and the main results also hold for QPSK).

In the first phase, we assume full channel information at the relay node (receiver node) and no channel information at the end nodes (transmitter nodes). In particular, the effects of transmit power and carrier-phase difference are combined into the complex channel coefficients. Eq (1) can be rewrite in the vector form as

$$R = HX + N \qquad (2)$$

Throughout this paper, a capital letter, for example $H$, denotes a matrix or a vector and the corresponding lower case letter, for example $h_{i,j}$, denotes its element on $i$-th row and $j$-th column. The relay node then tries to estimate the network coded form of the two end nodes' signals (i.e., $x_1 \oplus x_2$ in this paper).

In the second phase, the relay node broadcasts the estimated network-coded packet to both end nodes. Each end node then decodes its own target packet from the received network-coded packet with self information. The second phase is the same as a traditional MIMO broadcast with standard network decoding [13]. This paper focuses on the first phase.

**B. Illustrating Example:**

For the first phase, the transmission in (2) can be regarded as a point-to-point 2-by-2 MIMO system (a distributed MIMO system). The goal of the relay node is to obtain an estimate of $x_1 \oplus x_2$. In a traditional MIMO NC scheme, the processing of the relay node is to explicitly decode $x_1$ and $x_2$ before network-encoding them into $x_1 \oplus x_2$. However, this scheme is suboptimal since it does not make use of the fact that only $x_1 \oplus x_2$ rather than individual $x_1$ and $x_2$ is needed at the relay node. We now present an example to illustrate the suboptimality of the MIMO NC processing. This example also reveals the advantages of our proposed scheme.

Consider a special scenario in satellite communication where the two end nodes are on the earth and the relay node is the satellite. Suppose it is a line-of-sight channel without any multipath and the two end nodes' signal arrive at the two antennas of the satellite in a synchronous way. Nevertheless, the channel matrix could still be realized in many forms. For example, it could be

$$H = \begin{bmatrix} 1 & 1 \\ 1 & 1 \end{bmatrix}. \qquad (3)$$

In (3), $H$ is not a full-rank matrix and the relay can never obtain $x_1$ and $x_2$ individually from the received signal vector $R$. As a result, the multiple access rate of $x_1 \oplus x_2$, based on the MIMO NC scheme, is zero.

With PNC, the goal of the relay is to estimate $x_1 \oplus x_2$ from $R$ without first estimating $x_1$ and $x_2$. In fact, the information on $x_1+x_2$, which can be obtained directly from $R$ by matrix multiplication, is a more useful intermediate step as far as the estimation of $x_1 \oplus x_2$ is concerned.

Based on the above observation, we propose the following MIMO PNC scheme. We now illustrate the basic idea based on the specific $H$ in (3). The treatment for general $H$ can be found in Section III. With the $H$ in (3), the relay first combines the signals from the two receiving antennas as

$$r = \frac{1}{2}(r_1 + r_2) = x_1 + x_2 + \frac{n_1 + n_2}{2}. \qquad (4)$$

After that, the relay maps $r$ to $x_1 \oplus x_2$ according to the PNC mapping in [2]. As a result, in this new scheme, the relay can obtain $x_1 \oplus x_2$ with almost full rate [8-10]. This example shows that the proposed scheme may outperform the MIMO NC scheme significantly. In the following sections, we elaborate our proposed scheme, which makes use of linear MIMO detection. We prove that it outperforms the traditional MIMO NC scheme for all channel realizations of $H$.

## III. MIMO PNC DETECTION SCHEME

In this section, we present the details of our proposed detection and encoding scheme to obtain $x_1 \oplus x_2$ from the received signals, inspired by the basic idea of PNC [2]. Before that, we first review the traditional two phase relay scheme based on linear MIMO detection approaches for a purpose of comparison, i.e., the MIMO NC scheme.

### A. Detection and Encoding Based on Linear MIMO Detection:

As mentioned in the previous section, the objective of the relay is to obtain the XOR of the two end nodes information, i.e., $x_1 \oplus x_2$. A major detection method in MIMO for spatial-multiplexed systems is linear detection followed by quantization (e.g., [2]).

Let us first consider the application of this method on the traditional MIMO NC scheme in which $x_1$ and $x_2$ are to be explicitly estimated. First, an estimate of the transmitted information is calculated by multiplying an equalization matrix $G$ to both sides of (2) as

$$Y = GR = GHX + GN. \qquad (5)$$

After that, the detected data is obtained by componently quantizing $Y$ according to the symbol alphabet used (the alphabet is $\{1,-1\}$ for BPSK modulation). Specifically, the quantitative estimate of $X$ is

$$\tilde{x}_i = \begin{cases} 1 & \text{when } y_i \geq 0 \\ -1 & \text{when } y_i < 0 \end{cases}. \qquad (6)$$

At last, the estimates of $x_1$, $x_2$ in (6) are combined to obtain the network-coded symbol:

$$\widetilde{(x_1 \oplus x_2)} = (\tilde{x}_1) \oplus (\tilde{x}_2). \qquad (7)$$

The *Zero-Forcing* (ZF) equalizer is given by setting $G$ to the pseudo-inverse [14] of $H$, i.e., $G = H^+ = (H^H H)^{-1} H^H$. Zero forcing has very low complexity. However, it performs poorly when the condition number of $H$ is large. The *minimum mean square error* (MMSE) equalizer is given by [15], where $G$ is set to $G = (\sigma^2 I + H^H H)^{-1} H^H$. MMSE estimation minimizes the mean-square error $E\{\|Y - X\|^2\}$. Generally speaking, MMSE can outperform ZF, but with the cost of higher complexity and the requirement for additional information, i.e., the noise variance.

### B. Proposed MIMO PNC Scheme Based on Linear Detection:

This part presents our proposed MIMO PNC scheme. In this scheme, the relay node first obtains an estimate of $x_1+x_2$ and $x_1-x_2$, rather than individual $x_1$ and $x_2$, from the received signal. After that, it transforms both $x_1+x_2$ and $x_1-x_2$ to the target signal $x_1 \oplus x_2$ with PNC mapping. Let us consider the zero forcing (ZF) detection as an example to elaborate the details of the scheme.

The received signal in (2) can be re-written in the following form:

$$R = HX + N = (HD^{-1})(DX) + N = \hat{H}\hat{X} + N \qquad (8)$$

where $D = 2D^{-1} = \begin{bmatrix} 1 & 1 \\ 1 & -1 \end{bmatrix}$ is referred as the sum-difference matrix. For linear detection, we can similarly find the equalization matrix $G$ corresponding to $\hat{H}$ to calculate the estimate of $\hat{X}$ as in (5).

For ZF detection, $G = (\hat{H}^H \hat{H})^{-1} \hat{H}^H$ is the Moore-Penrose pseudo inverse of $\hat{H}$, and the estimate of $\hat{X}$ is $Y = GR$. Note that

$$\hat{X} = \begin{pmatrix} \hat{x}_1 \\ \hat{x}_2 \end{pmatrix} = \begin{pmatrix} x_1 + x_2 \\ x_1 - x_2 \end{pmatrix}. \qquad (9)$$

Obviously, $\hat{x}_1$ and $\hat{x}_2$ are correlated with each other and each of them can be mapped to $x_1 \oplus x_2$ with PNC mapping. We should combine the information from both $y_1$ and $y_2$ to obtain the estimate of the target signal $x_1 \oplus x_2$.

Due to the distinction between $\hat{x}_1$ and $\hat{x}_2$, we can not apply the maximum ratio combination, which is known to be optimal in maximizing SNR. As an alternative, we derive the Likelihood Ratio (LR) of $x_1 \oplus x_2$ from both $y_1$ and $y_2$. Ignoring the dependences between the noises in $y_1$ an $y_2$ as in conventional ZF processing, the likelihood ratio of $x_1 \oplus x_2$ can be written as

$$\begin{aligned} L(x_1 \oplus x_2 \mid y_1 y_2) &= \frac{P(y_1 y_2 \mid x_1 \oplus x_2 = 1)}{P(y_1 y_2 \mid x_1 \oplus x_2 = -1)} \\ &= \frac{[P(y_1 \mid \hat{x}_1 = 2) + P(y_1 \mid \hat{x}_1 = -2)]P(y_2 \mid \hat{x}_2 = 0)}{P(y_1 \mid \hat{x}_1 = 0)[P(y_2 \mid \hat{x}_2 = 2) + P(y_2 \mid \hat{x}_2 = -2)]} \\ &= L(x_1 \oplus x_2 \mid y_1) L(x_1 \oplus x_2 \mid y_2) \\ &= \exp(2/\sigma_2^2 - 2/\sigma_1^2) \cosh(2y_1/\sigma_1^2)/\cosh(2y_2/\sigma_2^2) \end{aligned} \qquad (10)$$

where $\sigma_i^2 = \{G^H G\}_{i,i} \sigma^2$ is the variance of the noise on the $i$-th stream after the zero-forcing signals de-mix. The corresponding decision rule should be

$$\widetilde{x_1 \oplus x_2} = \begin{cases} 1 & \text{when } L(x_1 \oplus x_2 \mid y_1 y_2) \geq 1 \\ -1 & \text{when } L(x_1 \oplus x_2 \mid y_1 y_2) < 1 \end{cases}. \qquad (11)$$

Eq. (10) shows that the Log Likelihood Ratio (log value of the LR in (10)) of the target signal is the summation of the LLR of each data stream and we refer to the combination in (10) as the LLR combination.

Although the LLR combination performs best, it needs more calculation and extra information, such as the Gaussian noise variance. We now consider the simple selective combination scheme in which one of $y_1$ or $y_2$ is chosen for our decision making, depending on the relative magnitude of the noises in $y_1$ and $y_2$. Specifically,

$$\widetilde{x_1 \oplus x_2} = \begin{cases} sign(abs(y_1) - thr) & \text{when } \{GG^H\}_{1,1} < \{GG^H\}_{2,2} \\ sign(thr - abs(y_2)) & \text{otherwise} \end{cases}. \quad (12)$$

where the sign function returns the sign of its parameter and the optimal threshold *thr* in (12) can be obtained as in [2], or we could simply set *thr*=1 in high SNR region with little performance loss.

For another popular linear detection scheme, MMSE detection, $G = (\hat{H}^H \hat{H} + \sigma^2 I_2)^{-1} \hat{H}^H$. And the corresponding LLR based and selective based decision rules can be obtained similarly and they are omitted here due to limited space.

### C. BER Performance Analysis

This part analyzes the BER performance of the ZF-based MIMO PNC scheme and compares it with the two-phase MIMO NC scheme in which Zero Forcing MIMO detection method is used to extract the individual packets from the two end nodes before converting them into a network-coded packet. We first introduce the following conjecture and lemma.

In the ZF-based MIMO PNC scheme, assume that the sum of the variances of the two data streams after data de-mix is a constant (i.e., $\sigma_1^2 + \sigma_2^2 = (G_{11}^2 + G_{12}^2 + G_{21}^2 + G_{22}^2)\sigma^2 = c$), and without loss of generality, assume $\sigma_1^2 \leq \sigma_2^2$. Then the PNC mapping based only on $y_1$ in fact correspond to non-MIMO PNC processing. The associated BER of $x_1 \oplus x_2$ is very close to a point-to-point transmission system with noise variance $\sigma_1^2$. As the difference between $\sigma_1^2$ and $\sigma_2^2$ increases (i.e., $\sigma_1^2$ decreases and $\sigma_2^2$ increases while keeping $\sigma_1^2 + \sigma_2^2 = c$), the BER of PNC mapping based only on $y_1$ decreases. This BER, on the other hand, serves as an upper bound of the BER resulting from our decision rule in (10) since the combination method in (10) makes use of both $y_1$ and $y_2$. Because the upper bound of the BER of the ZF-based MIMO-PNC scheme is maximized when $\sigma_1^2 = \sigma_2^2$, we make the following conjecture that the BER itself is also maximized when $\sigma_1^2 = \sigma_2^2$ (note: this conjecture has been verified by numerical results from simulation):

*Conjecture 1*: Consider the ZF-based MIMO PNC scheme that makes use of the decision rule in (10). Suppose that the sum of the variances of the two data streams after de-mixing is a constant (i.e., $\sigma_1^2 + \sigma_2^2 = c$). The BER is maximized when $\sigma_1^2 = \sigma_2^2$.

To explain the next lemma, consider a special case where the channel matrix is

$$H = \begin{bmatrix} h_{1,1} & 0 \\ 0 & h_{2,2} \end{bmatrix}. \quad (13)$$

where $h_{1,1}$ and $h_{2,2}$ are random complex channel coefficients. Then the two received signals at the relay can be expressed as follows by equalizing the channel effect:

$$z_1 = x_1 + n_1 \qquad z_2 = x_2 + n_2 \quad (14)$$

where the noise variances of the two data streams are $\sigma_1^2 = \sigma^2 / |h_{1,1}|^2$, $\sigma_2^2 = \sigma^2 / |h_{2,2}|^2$. With the detection and encoding method in (7), suppose that $x_1 \oplus x_2$ can be obtained with a BER denoted by $P_1(\sigma_1^2, \sigma_2^2)$. Superimposing and subtracting the two signals in (14), we can obtain

$$y_1 = z_1 + z_2 \qquad y_2 = z_1 - z_2. \quad (15)$$

The variances of noises in both $y_1$ and $y_2$ are both $\sigma_1^2 + \sigma_2^2$. With the detection and encoding method in (10) and (11), suppose that $x_1 \oplus x_2$ can be obtained with a BER denoted by $P_2(\sigma_1^2 + \sigma_2^2)$. Then we have the following lemma:

*Lemma 2*: For the special channel in (13), we always have that

$$P_1(\sigma_1^2, \sigma_2^2) = P_2(\sigma_1^2 + \sigma_2^2). \quad (16)$$

This lemma can be proved by comparing the noise region of both schemes in (14) and (15), where an error occurs. The details of the proof are omitted due to the limited space. Intuitively, this result is due to the independence of $n_1$ and $n_2$ in (14), which results in the same BER for the two optimal linear processings as in (14) and (15).

Based on *Conjecture 1* and *Lemma 2*, we have the following proposition for the BER performance of the ZF-based MIMO PNC scheme.

*Proposition 3*: For any given channel *H*, the BER of the proposed ZF-based MIMO PNC scheme is always no worse than the BER of the MIMO NC scheme, if *Conjecture 1* is true.

Proof: Let us first discuss the BER based on traditional MIMO NC detection and encoding scheme, which is denoted by $P_{tra}$. After ZF de-mixing as in (6), the noise variance of the *i*-th (*i*=1 or 2) data stream is

$$\sigma_i^2 = \left[ (H^H H)^{-1} \right]_{i,i} \sigma^2 = \Sigma_{i,i} \sigma^2 \quad (17)$$

In (17), *H* is the channel matrix. For any *H*, $\Sigma = (H^H H)^{-1}$ is an Hermitian Matrix and it can be decomposed with singular value decomposition as

$$\Sigma = \begin{bmatrix} \cos(\alpha) & \sin(\alpha) \\ \sin(\alpha) & -\cos(\alpha) \end{bmatrix} \begin{bmatrix} \mu_1 & 0 \\ 0 & \mu_2 \end{bmatrix} \begin{bmatrix} \cos(\alpha) & \sin(\alpha) \\ \sin(\alpha) & -\cos(\alpha) \end{bmatrix}^H \quad (18)$$

where $\mu_1$, $\mu_2$ and $\alpha$ are real values. Then we have

$$\Sigma_{1,1} = \cos^2(\alpha)\mu_1 + \sin^2(\alpha)\mu_2 \\ \Sigma_{2,2} = \cos^2(\alpha)\mu_2 + \sin^2(\alpha)\mu_1. \quad (19)$$

According to the method in (7), the BER of $x_1 \oplus x_2$ only depends on the variances of the two noises in (19), rather than the covariance between them. With the notation in Lemma 2, $P_{tra}$ can be expressed as

$$P_{tra} = P_1(\Sigma_{1,1}\sigma^2, \Sigma_{2,2}\sigma^2). \quad (20)$$

Based on Lemma 2, we can further obtain that

$$P_{tra} = P_2(\sigma^2(\Sigma_{1,1} + \Sigma_{2,2})) = P_2(\sigma^2(\mu_1 + \mu_2)). \quad (21)$$

Let us then discuss the BER of the MIMO PNC detection, which is denoted by $P_{MIMO\ PNC}$. Based on the data de-mix method in (9), the variance of the noise in $y_i$ is

$$\hat{\sigma}_i^2 = \left[(\hat{H}^H \hat{H})^{-1}\right]_{i,i} \sigma^2 = \hat{\Sigma}_{i,i}\sigma^2. \quad (22)$$

Here $\hat{\Sigma} = (\hat{H}^H \hat{H})^{-1} = 2D^{-1}\sum D$ and it can be decomposed as

$$\hat{\Sigma} = 2\begin{bmatrix}\cos(\beta) & \sin(\beta) \\ \sin(\beta) & -\cos(\beta)\end{bmatrix}\begin{bmatrix}\mu_1 & 0 \\ 0 & \mu_2\end{bmatrix}\begin{bmatrix}\cos(\beta) & \sin(\beta) \\ \sin(\beta) & -\cos(\beta)\end{bmatrix}^H. \quad (23)$$

Similar to (19), we have $\hat{\sigma}_1^2 + \hat{\sigma}_2^2 = 2\sigma^2(\mu_1 + \mu_2)$. With fixed $(\mu_1 + \mu_2)$, the worst BER is achieved when $\hat{\sigma}_1^2 = \hat{\sigma}_2^2 = \sigma^2(\mu_1 + \mu_2)$ ($\beta = \pi/4$ or $\mu_1 = \mu_2$) according to *Conjecture 1*. And this BER, which can be expressed as $P_2(\sqrt{\mu_1 + \mu_2})$ as in the special case, is no less than $P_{MIMO\ PNC}$. Therefore, we prove our proposition as

$$P_{MIMO\_PNC} \leq P_2(\sqrt{\mu_1 + \mu_2}) = P_{tra}. \quad (24)$$

An intuitive explanation of this proposition is as follows. Any channel matrix $H = [h_{i,j}]$ can be regarded as the summation of two sub-matrixes as

$$H = \begin{bmatrix}h_{1,1} & h_{1,2} \\ h_{2,1} & h_{2,2}\end{bmatrix} = \begin{bmatrix}h_{1,1} - h_{1,2} & 0 \\ 0 & h_{2,2} - h_{2,1}\end{bmatrix} + \begin{bmatrix}h_{1,2} & h_{1,2} \\ h_{2,1} & h_{2,1}\end{bmatrix}. \quad (25)$$
$$= H_1 + H_2$$

For $H_1$, the BER of the two schemes is the same as discussed above. For $H_2$, the BER performance of the MIMO NC scheme is always 0.5 while the BER of the MIMO PNC scheme is much smaller (which depends on the relative SNR). As a result, our MIMO PNC scheme outperforms the traditional scheme for all channel realizations.

## IV. SIMULATION AND EXTENSION

In this section, we first present some numerical simulation results for MIMO PNC. After that, we discuss some extensions of this scheme.

### A. Numerical Simulation:

The simulation setting is mainly based on the system model in *Section I*. The variance of the complex channel coefficient is set to 1 on each dimension and the SNR of the system is defined as $1/\sigma^2$. The simulation focuses on the BER of $x_1 \oplus x_2$ at the relay node since the broadcast phase is the same as that in traditional MIMO broadcast system.

In Figure 2, we plot the BER of the proposed ZF-based MIMO PNC schemes (LLR based decisions in (11) and selective based decisions in (12)) are plotted under random complex channel matrix. We also plot the BER of the ZF-based MIMO NC scheme (7) for comparison. As shown in this figure, the proposed scheme with LLR combination outperforms the traditional scheme by about 1.6dB. When the BER is less than 1$e$-2, the proposed scheme with selective combination outperforms the traditional scheme by about 1 dB. Note that this improvement is achieved without any extra cost.

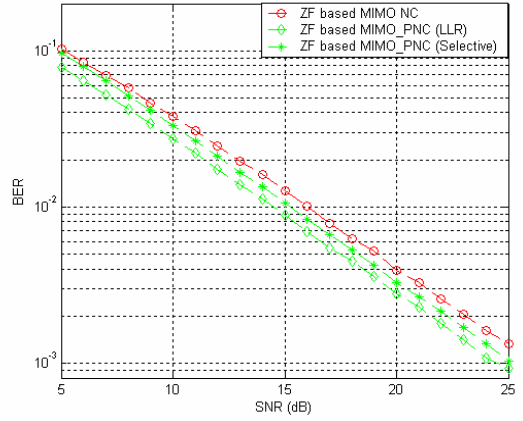

Figure 2. BER performance of the ZF based MIMO PNC schemes and the traditional ZF scheme

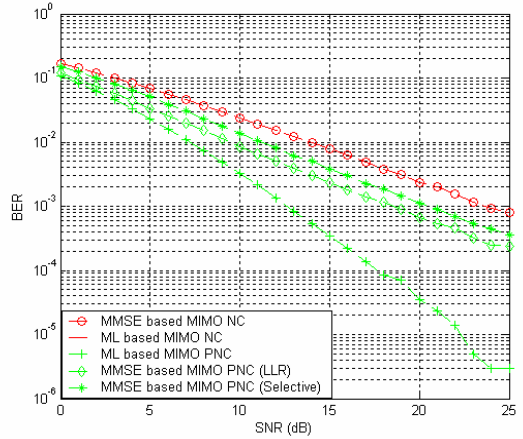

Figure 3. BER performance of the MMSE based MIMO PNC schemes and the MIMO NC scheme

In Figure 3, the BER of the MMSE-based traditional MIMO scheme and the BER of the proposed MMSE -based MIMO PNC schemes are plotted under random complex channel matrix. We can see that the proposed scheme with LLR combination outperform the traditional scheme by about 5.5dB when the BER is less than 1$e$-3, while the proposed scheme with selective combination outperforms the traditional scheme by about 3.5 dB. This significant performance improvement is of more interest by noting that the MMSE based MIMO detection schemes are widely used in current wireless systems. Figure 3 also shows the BER at the relay with the optimal maximum likelihood (ML) detection and encoding schemes[1]. As shown in the figure,

---
[1] For the ML based MIMO NC, $(\tilde{x}_1, \tilde{x}_2) = \arg\max_{(x_1, x_2)} \Pr(y_1, y_2 \mid x_1, x_2)$ and

when SNR is less than 5dB, the proposed MMSE based MIMO PNC scheme (LLR combination) performs close to the optimal ML scheme.

In fact, the proposed MMSE based MIMO PNC scheme improves the diversity from 1 to 2 in low SNR region. The explanation is similar to the discussion in [16] and the rigorous proof is future work.

### B. Discussion

In order to illustrate the basic idea of MIMO PNC, we assume two antennas at the relay node and one antenna at each end node. As a result, the sum-difference matrix $D$ is a 2-by-2 matrix. When there are more than two antennas at the relay node, such a 2-by-2 sum-difference matrix $D$ is still workable. Consider a more general scenario that there are $L$ antennas at each end nodes and there are more than $L$ antennas at the relay node. Denoting the data transmitted on the $i$-th antenna of the two end nodes by $x_i$, $y_i$ respectively, the $2L$-by-$2L$ sum-difference matrix could be

$$D_{2L} = \begin{bmatrix} D & 0 & 0 \\ 0 & \ddots & 0 \\ 0 & 0 & D \end{bmatrix} \quad (26)$$

where the end nodes' data are listed as a column vector $[x_1, y_1, x_2, y_2, \cdots x_L, y_L]^T$.

The essential idea of MIMO PNC is to find a matrix (the sum-difference matrix) which satisfies the following two conditions: i) this matrix matches the wireless channel so that the linear transformation from the received signal to the mixed form with the sum-difference matrix loses little information; ii) original signals $(x_1, x_2)$ mixed with the sum-difference matrix can be easily transformed to their network coding form with little information loss. Therefore, the optimal sum-difference matrix may depend on the given channel matrix. When considering fast fading wireless channels where the channel matrix changes from symbol to symbol, the sum-difference matrix that we choose in (8) is favorable since it is independent of the channel matrix.

In order to calculate the uncoded BER, the estimate of $x_1 \oplus x_2$ at the relay is hard decided in our paper. In fact, the estimate can be easily transformed to the soft version for the ease of soft channel decoding and the ease of soft forwarding.

## V. CONCLUSION

In this paper, a novel signal detection and network encoding scheme, MIMO PNC, is proposed to extract $x_1 \oplus x_2$ from the superimposed signals received at the multiple antennas of the relay node. Different from the traditional MIMO NC scheme where the relay tries to obtain individual $x_1$ and $x_2$ with standard MIMO detection methods before converting them into $x_1 \oplus x_2$, our new scheme first tries to obtain $x_1$-$x_2$ and $x_1$+$x_2$ with linear MIMO detection methods at the relay before converting them to $x_1 \oplus x_2$ with PNC mapping. As shown in our illustrating example, this simple scheme can effectively improve the performance. Further analysis shows that our ZF based MIMO-PNC scheme may always outperform the traditional ZF based MIMO NC scheme for any given channel matrix. The simulation results verify the advantages of our new schemes under the setting of random Rayleigh fading channel coefficients. In particular, a SNR improvement of 5.5 dB can be observed for the widely used MMSE based detection schemes.

ACKNOWLEDGMENT

This project is supported by the National Science Foundation of China (Grant No. 60902016) and RGC grant CERG 414507.

---

$\widetilde{(x_1 \oplus x_2)} = (\tilde{x}_1) \oplus (\tilde{x}_2)$. For the ML based MIMO PNC, the decision rule is $\widetilde{(x_1 \oplus x_2)} = \arg\max_{\pm 1} \Pr(y_1, y_2 | x_1 \oplus x_2)$. As shown in Fig. 3, the two schemes perform very close to each other.